%% file: main.tex
\pdfoutput=1
\documentclass[letterpaper]{article}
\usepackage{spconf}
\providecommand{\IEEEauthorblockN}[1]{#1}
\providecommand{\IEEEauthorblockA}[1]{#1}
\usepackage{amsmath,amssymb,graphicx,booktabs}
\usepackage{tikz}
\usetikzlibrary{arrows.meta,positioning,calc}
\usepackage[hidelinks]{hyperref}
\title{Reward-Tilted On-Policy Distillation for Acoustic Grounding in Audio-Language Models}

\name{\IEEEauthorblockN{Kaiyang Li\textsuperscript{1,2 $^\dagger$}\thanks{This work was conducted during an internship at NEC Labs.}, Shaobo Han\textsuperscript{1}, Yue Tian\textsuperscript{1}, Shihao Ji\textsuperscript{2}}}
\address{\IEEEauthorblockA{
        \textsuperscript{1}NEC Laboratories America, Inc., USA \\
        \textsuperscript{2}School of Computing, University of Connecticut, USA \\
    }}

\makeatletter
\renewcommand\section{\@startsection{section}{1}{\z@}{-9pt plus -2pt minus -1pt}{4pt plus 1pt}{\normalfont\normalsize\bfseries}}
\renewcommand\subsection{\@startsection{subsection}{2}{\z@}{-8pt plus -2pt minus -1pt}{3pt plus 1pt}{\normalfont\normalsize\bfseries}}
\makeatother
\makeatletter
\long\def\@makefntext#1{%
  \parindent\z@\hangindent1.2em\hangafter\@ne
  \noindent\hb@xt@1.2em{\hss
    \ifx\@thefnmark\@empty\textsuperscript{\ensuremath{\dagger}}%
    \else\textsuperscript{\@thefnmark}\fi\hskip.35em}#1}
\makeatother

\begin{document}
\maketitle

\begin{abstract}

Audio-language models (ALMs) can exploit textual shortcuts to answer questions while overlooking acoustic evidence, weakening audio understanding. On-policy distillation (OPD) trains compact ALMs by supervising student-generated responses with teacher predictions, but does not explicitly distinguish acoustic support from linguistic predictability. We propose Reward-Tilted On-Policy Distillation (RT-OPD) to strengthen acoustic grounding. Given the same question and student-generated text, a frozen teacher predicts the next token with and without audio inputs. Their log-probability contrast defines a reward that reshapes the teacher distribution for reverse-KL distillation, emphasizing the additional evidence provided by audio. Across two compact students and three benchmarks, RT-OPD consistently outperforms Vanilla OPD. Experiments with silenced and replacement audio further suggest that RT-OPD strengthens the student's reliance on acoustic evidence. Our 3B model achieves 72.72\% accuracy on MMAU, the highest among the compared 3B models and competitive with several 7B and 8B models. Code and model checkpoints are available at \url{https://github.com/KaiyangLi1992/RT-OPD}.

\end{abstract}


\nocite{ref2,ref3,ref4,ref7,ref8,he2025audiomcq,ref20}

\begin{table}[t]

\centering

\caption{ALM performance on MMAU (\%), ranked by accuracy. External scores and reported sizes follow the official leaderboard~\cite{ref1}. Mizar-3B\protect\footnotemark[1] is obtained by post-training Ke-Omni-R-3B with RT-OPD; its score is averaged over five seeds. Protocols and parameter-count conventions differ across models.}

\vspace{3pt}

\begin{tabular}{@{}lrr@{}}

\toprule

Model & Reported size & MMAU\\

\midrule

Audio-Thinker~\cite{audio_thinker} & 8.4B & 75.98\\

Nova 2 Omni~\cite{nova2} & --- & 75.28\\

Step-Audio-2~\cite{stepaudio2} & --- & 73.86\\

\textbf{Mizar-3B (ours)} & \textbf{3B} & \textbf{72.72}\\

MiMo-Audio~\cite{mimoaudio} & 7B & 72.59\\

Audio Flamingo 3~\cite{af3} & 8.2B & 72.42\\

Qwen2.5-Omni~\cite{ref4} & 8.2B & 71.00\\

Gemini 2.5 Pro~\cite{gemini25} & --- & 69.36\\

Audio Flamingo 2 Reasoning~\cite{ref1} & 3B & 64.70\\

Kimi-Audio~\cite{kimiaudio} & 8.2B & 64.40\\

Audio Flamingo 2~\cite{af2} & 3B & 61.06\\

GPT-4o Audio~\cite{gpt4o} & --- & 60.82\\

\bottomrule

\end{tabular}

\label{tab:1}
\end{table}
\footnotetext[1]{The name ``Mizar'' was inspired by the star used for navigation. We hope our model and method can provide guidance for future research and engineering optimization on audio-language models.}

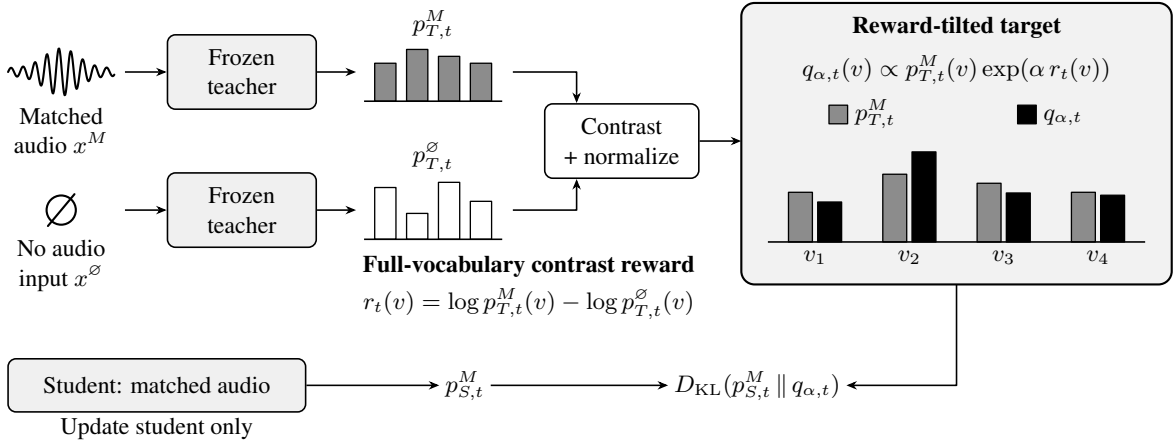
\begin{figure*}[t]

\centering

\input{figures/figure1_tikz.tex}

\caption{\textbf{Overview of RT-OPD.} Given the same question and student-generated text, a frozen teacher predicts the next token with and without audio. Their log-probability difference is a reward that reshapes the teacher distribution: candidates with larger audio-present/\allowbreak absent probability ratios are boosted more before normalization. The student learns from this target through reverse KL. Bars are schematic; gray, white, and black bars depict $p^{M}_{T,t}$, $p^{\varnothing}_{T,t}$, and $q_{\alpha,t}$, respectively.}

\label{fig:1}
\end{figure*}

\section{Introduction}

Audio-language models (ALMs) answer open-ended questions about speech, environmental sound, and music by combining acoustic perception with language understanding and reasoning~\cite{ref2,ref3,ref4}. Deploying these models under limited memory and computation requires compact ALMs that retain strong reasoning capabilities. On-policy distillation (OPD) enables transferring reasoning capabilities from larger ALMs to compact students by supervising their own generated responses, reducing exposure bias through teacher guidance on student-generated prefixes~\cite{ref7,ref8}.

Recent studies reveal that ALMs could exploit textual shortcuts to answer audio questions without relying on the recordings~\cite{he2025audiomcq,ref20}. Such shortcuts can mask weaknesses in audio understanding. For example, a model may complete ``a dog is  \underline{\hspace{0.5cm}}'' with ``barking'' based on linguistic context alone; producing ``a dog is barking'' does not establish that it has identified ``barking'' in the recording. However, standard OPD matches the teacher's audio-conditioned distribution without explicitly emphasizing the additional evidence provided by the recording.

We propose \emph{Reward-Tilted On-Policy Distillation} (RT-OPD) to emphasize acoustic evidence in the teacher's distillation targets. As shown in Fig.~\ref{fig:1}, given the same question and student-generated text, a frozen teacher predicts the next token with and without audio. We define the reward as the log-probability difference between the audio-present and audio-absent conditions. That is, a token receives a larger reward when its probability with audio is higher than its probability without audio. We use this reward to reweight the original teacher distribution, giving greater weight to tokens with larger rewards. The reweighted probabilities are then normalized to form a new distillation target, encouraging the student to learn predictions supported by acoustic evidence.

We evaluate RT-OPD with Ke-Omni-R-3B~\cite{ref9} and Qwen2.5-Omni-3B~\cite{ref4} students on MMAU~\cite{ref18}, MMAR~\cite{ref19}, and ADQA-cl.~\cite{ref20}. Averaged over five seeds, RT-OPD consistently outperforms Vanilla OPD across all three benchmarks for both students. Further analysis suggests that RT-OPD relies more strongly on acoustic evidence than Vanilla OPD. Mizar-3B, our model obtained by post-training Ke-Omni-R-3B with RT-OPD, reaches 72.72\% accuracy on MMAU, the highest among the compared 3B ALMs and competitive with several 7B and 8B models (Table~\ref{tab:1}).

\section{Related Work}

\textbf{Audio Reasoning and Distillation.}
Recent work improves audio-language reasoning through reinforcement learning and knowledge distillation. Ke-Omni-R~\cite{ref9} and CESAR~\cite{ref10} use GRPO to strengthen the chain-of-thought reasoning in audio-language models. Distillation-based approaches instead transfer reasoning behavior from stronger teachers to compact models. CORD~\cite{ref11} combines weighted reverse-KL alignment to a teacher with sequence-level GRPO, while X$^3$-OPD~\cite{ref12} performs reasoning distillation along trajectories generated by the student itself. These methods primarily focus on transferring or optimizing reasoning capability. RT-OPD instead focuses on how teacher supervision can explicitly emphasize predictions that are supported by acoustic evidence.

\textbf{Audio-Aware Contrastive Supervision.}
A related line of work compares model predictions under different audio conditions. Audio-aware decoding~\cite{ref13} contrasts predictions with and without audio at inference time to reduce audio hallucination. CAAD~\cite{ref14} brings the same type of audio-present/absent contrast into distillation, but trains on responses generated in advance by the teacher. In contrast, RT-OPD applies the contrast along student-generated trajectories and uses it to reshape the teacher's next-token distribution before reverse-KL distillation. Visual-Advantage OPD~\cite{ref15} also introduces modality-dependent contrast into on-policy distillation, but uses the contrast to weight trajectories or token groups rather than directly reshape vocabulary-level teacher predictions. G-OPD~\cite{ref16} independently derives a mathematically related target-shaping form using the contrast between a teacher and a separate reference model. RT-OPD instead compares the same frozen teacher with and without audio, so its contrast directly measures how acoustic evidence changes the teacher's support for each candidate token.

\section{Method}

\subsection{Standard OPD and Setup}

Given an audio input $x^M$ and a question $a$, the student generates a response $y=(y_1,\ldots,y_L)$ of length $L$. At each position $t$, let $\xi_t^M=(x^M,a,y_{<t})$ denote the context consisting of audio, question, and student-generated prefix. Conditioned on this context, the teacher and the student predict the next-token distributions $p_{T,t}^{M}(v)=p_T(v\mid\xi_t^M)$ and $p_{S,t}^{M}(v)=p_S(v\mid\xi_t^M)$, respectively, where $v\in\mathcal{V}$ and $\mathcal{V}$ is the vocabulary. Standard OPD trains the student to match the teacher along the generated response by minimizing\vspace{-2pt}
\begin{equation}
\mathcal{L}_{\mathrm{OPD}}=\frac{1}{L}\sum_{t=1}^{L}D_{\mathrm{KL}}\!\left(p_{S,t}^{M}\,\|\,p_{T,t}^{M}\right).
\tag{1}\vspace{-1pt}
\end{equation}

\subsection{Acoustically Grounded Teacher-Target Shaping}

To measure how audio changes the teacher's support, we evaluate the same frozen teacher with the audio input removed while retaining the question and student-generated text. We denote this reference context by $\xi_t^{\varnothing}=(x^{\varnothing},a,y_{<t})$ and its teacher distribution by $p_{T,t}^{\varnothing}(v)=p_T(v\mid\xi_t^{\varnothing})$. The log-probability contrast defines a reward for every vocabulary candidate:
\begin{equation}
r_t(v)=\log p_{T,t}^{M}(v)-\log p_{T,t}^{\varnothing}(v).
\tag{2}
\end{equation}

A larger reward indicates a larger proportional increase in teacher support when audio is supplied. We use this signal to reweight the original teacher distribution and normalize it into a distillation target:
\begin{equation}
q_{\alpha,t}(v)=
\frac{p_{T,t}^{M}(v)\exp\!\left(\alpha r_t(v)\right)}
{\sum_{u\in\mathcal{V}}p_{T,t}^{M}(u)\exp\!\left(\alpha r_t(u)\right)},
\quad \alpha\geq0.
\tag{3}
\end{equation}
The coefficient $\alpha$ controls the adjustment strength; in particular, $\alpha\!=\!0$ recovers standard OPD. Combining Eqs. (2) and (3), the target can also be written as
\begin{equation*}
q_{\alpha,t}(v)\propto\frac{p_{T,t}^{M}(v)^{1+\alpha}}{p_{T,t}^{\varnothing}(v)^{\alpha}}.
\end{equation*}
When $\alpha\!>\!0$, candidates with larger audio-present/\allowbreak absent probability ratios receive stronger amplification relative to other candidates, directing student learning toward predictions that gain support from audio.

\subsection{Student Training Objective}

The student learns from the reshaped target under the original audio. We average the distillation loss over the $L_i$ valid completion tokens in the $i$-th response:\vspace{-5pt}
\begin{equation}
\mathcal{L}_{\mathrm{RT},i}=\frac{1}{L_i}\sum_{t=1}^{L_i}
D_{\mathrm{KL}}\!\left(p_{S,t}^{M}\,\|\,q_{\alpha,t}\right).
\tag{4}\vspace{-2pt}
\end{equation}
Substituting the reshaped target from Eq.~(3) into the token-level KL term in Eq.~(4) yields
\begin{equation}
\begin{aligned}
D_{\mathrm{KL}}(p_{S,t}^{M}\|q_{\alpha,t})
&=D_{\mathrm{KL}}(p_{S,t}^{M}\|p_{T,t}^{M})\\
&\quad-\alpha\,\mathbb{E}_{v\sim p_{S,t}^{M}}[r_t(v)]+\log Z_t.
\end{aligned}
\tag{5}
\end{equation}
Eq.~(5) decomposes the distillation loss into three terms. The first term, $D_{\mathrm{KL}}(p_{S,t}^{M}\|p_{T,t}^{M})$, keeps the student close to the original teacher distribution. The second term, $-\alpha\allowbreak\mathbb{E}[r_t(v)]$, encourages the student to assign more probability to candidates with larger rewards $r_t(v)$, defined in Eq.~(2) as the teacher's log-probability difference between audio-present and audio-absent conditions. The third term, $\log Z_t$, normalizes the target, with $Z_t=\allowbreak\sum_{u\in\mathcal{V}}p_{T,t}^{M}(u)\exp[\alpha r_t(u)]$; it is constant at a fixed prefix and does not affect student gradients.

We additionally use a cross-entropy loss $\mathcal{L}_{\mathrm{ce},i}$ to train the student to predict the gold answer $y^*$. A precomputed gate $g_i\in\{0,1\}$ enables distillation only when the teacher answers correctly with the original audio, giving the combined objective
\begin{equation}
\mathcal{L}=\frac{1}{B}\sum_{i=1}^{B}
\left[\mathcal{L}_{\mathrm{ce},i}+\lambda g_i\mathcal{L}_{\mathrm{RT},i}\right],
\tag{6}
\end{equation}
where $B$ is the global batch size. When the gate excludes a training example from distillation, its cross-entropy loss remains active. In our implementation, we set $\lambda\!\!=\!\!0.25$ and $\alpha\!\!=\!\!1$ and train the student with rank-64 LoRA~\cite{ref17}. The audio-absent teacher is used only during training, and the student's LoRA weights are merged before deployment.

\section{Experiments}

\subsection{Experimental Setup}

We train Ke-Omni-R-3B and Qwen2.5-Omni-3B students~\cite{ref9,ref4} using a frozen Ke-Omni-R-7B teacher~\cite{ref9}. Both students are trained on the same fixed subset of 10,000 audio multiple-choice question-answering examples randomly selected from AudioMCQ-StrongAC-GeminiCoT~\cite{ref20}. RT-OPD training runs for two epochs (626 steps) with a global batch size of 32 and learning rates of $7.5\!\times\! 10^{-5}$ for Ke and $2.5\!\times\! 10^{-5}$ for Qwen. The student responses are sampled at temperature 1.0 with top-p = 0.95, top-k = 64, and a maximum of 96 new tokens.

The 1,000-example MMAU test-mini set~\cite{ref18} serves as the validation set for hyperparameter tuning. Evaluation covers the separate 9,000-example MMAU test set (hereafter MMAU)~\cite{ref18}, MMAR~\cite{ref19}, and a cleaned version of the ADQA development set~\cite{ref20}. For the latter, removing examples whose audio overlaps with MMAU test-mini leaves 1,577 examples, forming ADQA-clean (ADQA-cl.). The training, validation, and test partitions are disjoint at the audio level. Evaluation uses greedy decoding with a maximum of 256 new tokens. Macro-3 is the arithmetic mean of accuracy across the three benchmarks. For trained methods, we report mean accuracy over five seeds; sample standard deviations are reported only for Macro-3.

\subsection{Main Results}

Table~\ref{tab:2} compares RT-OPD with the students before distillation, CE-only training, Vanilla OPD, and CAAD~\cite{ref14}. CE-only training employs cross-entropy on gold answers without supervising reasoning text. Vanilla OPD~\cite{ref7,ref8} combines this loss with reverse-KL distillation from the original teacher distribution, whereas RT-OPD uses the reshaped teacher target. CAAD~\cite{ref14} distills audio-present/absent teacher targets on pre-generated teacher responses using forward KL without gold-answer CE.

\begingroup\setlength{\tabcolsep}{0pt}

\begin{table}[t]

\centering

\caption{Main results (\%). Trained methods: five-seed means, with Macro-3 as mean $\pm$ sample SD over seeds; teacher and students before distillation: single runs. Bold: best mean per student block.}

\vspace{3pt}

\begin{tabular*}{\columnwidth}{@{\extracolsep{\fill}}lrrrr@{}}

\toprule

Method & MMAU & MMAR & ADQA-cl. & Macro-3\\

\midrule

\multicolumn{5}{l}{\emph{Ke-Omni-R-7B (teacher)}}\\

Teacher & 73.86 & 61.70 & 54.85 & 63.47\\

\midrule

\multicolumn{5}{l}{\emph{Ke-Omni-R-3B}}\\

Student & 68.70 & 52.70 & 49.59 & 57.00\\

CE only & 71.32 & 58.60 & 54.41 & 61.44 $\pm$ 0.34\\

Vanilla OPD~\cite{ref7} & 72.21 & 57.78 & 56.02 & 62.00 $\pm$ 0.31\\

CAAD~\cite{ref14} & 70.14 & 56.04 & 55.47 & 60.55 $\pm$ 0.18\\

RT-OPD (ours) & \textbf{72.72} & \textbf{60.08} & \textbf{56.45} & \textbf{63.08} $\pm$ 0.36\\

\midrule

\multicolumn{5}{l}{\emph{Qwen2.5-Omni-3B}}\\

Student & 61.67 & 42.40 & 38.55 & 47.54\\

CE only & 69.01 & 52.20 & 52.69 & 57.97 $\pm$ 0.42\\

Vanilla OPD~\cite{ref7} & 71.66 & 55.90 & 54.83 & 60.79 $\pm$ 0.24\\

CAAD~\cite{ref14} & 70.27 & 54.98 & 54.14 & 59.80 $\pm$ 0.36\\

RT-OPD (ours) & \textbf{72.18} & \textbf{58.90} & \textbf{55.70} & \textbf{62.26} $\pm$ 0.15\\

\bottomrule

\end{tabular*}

\label{tab:2}\vspace{-5pt}
\end{table}

\endgroup

RT-OPD outperforms both Vanilla OPD and CAAD on all three benchmarks for both students. Relative to Vanilla OPD, the largest gains occur on MMAR: 2.3 and 3.0 points for Ke and Qwen, respectively. These results support incorporating the teacher's audio-present/absent contrast into on-policy distillation.

Mizar-3B reaches a Macro-3 score of 63.08, close to the 7B teacher's 63.47, and exceeds the teacher on ADQA-clean. It also achieves the highest MMAU accuracy among the compared 3B models and remains competitive with several 7B and 8B models (Table~\ref{tab:1}).

\subsection{Ablation Study}
\vspace{-2pt}
\begingroup\setlength{\tabcolsep}{4.4pt}

\begin{table}[t]

\centering

\caption{Target and rollout ablations (\%). Macro-3 shows mean $\pm$ sample SD over per-seed scores. Bold marks the best mean within each student block.}

\vspace{3pt}

\begin{tabular}{@{}lrrrr@{}}

\toprule

Variant & MMAU & MMAR & ADQA-cl. & Macro-3\\

\midrule

\multicolumn{5}{l}{\emph{Ke-Omni-R-3B}}\\

RT-OPD & 72.72 & \textbf{60.08} & \textbf{56.45} & \textbf{63.08} $\pm$ 0.36\\

Unmatched & \textbf{72.84} & 59.06 & 56.09 & 62.66 $\pm$ 0.51\\

Sharpening & 72.59 & 56.95 & 54.57 & 61.37 $\pm$ 0.40\\

Model ref. & 72.57 & 58.10 & 53.44 & 61.37 $\pm$ 0.15\\

Off-policy & 71.72 & 58.84 & 56.17 & 62.24 $\pm$ 0.30\\

\midrule

\multicolumn{5}{l}{\emph{Qwen2.5-Omni-3B}}\\

RT-OPD & 72.18 & 58.90 & \textbf{55.70} & \textbf{62.26} $\pm$ 0.15\\

Unmatched & 72.26 & \textbf{58.94} & 55.23 & 62.15 $\pm$ 0.31\\

Sharpening & 71.77 & 56.10 & 55.16 & 61.01 $\pm$ 0.65\\

Model ref. & \textbf{72.31} & 57.83 & 54.11 & 61.42 $\pm$ 0.14\\

Off-policy & 70.96 & 57.68 & 55.08 & 61.24 $\pm$ 0.32\\

\bottomrule

\end{tabular}

\label{tab:3}\vspace{-5pt}
\end{table}

\endgroup

Table~\ref{tab:3} compares target and rollout variants. ``Unmatched'' uses the same teacher's predictions on a different recording of the same audio type as the reference, keeping the question and student-generated text fixed. ``Sharpening'' reshapes the original teacher distribution as $q_{\mathrm{sharp},t}=\allowbreak\operatorname{softmax}(z_{T,t}^{M}/\tau)$, where $z_{T,t}^{M}$ denotes the teacher logits under matched audio and $\tau=0.5$, without an audio contrast. ``Model reference'' uses a G-OPD-style contrast~\cite{ref16}: reference probabilities come from the frozen initial student under the original audio, question, and current-student-generated text, replacing the audio-absent teacher probabilities. ``Off-policy'' replaces online student responses with pre-generated teacher responses while retaining RT-OPD's audio-contrast target, reverse KL, and gold-answer CE.

RT-OPD achieves higher Macro-3 than ``Unmatched'' for both Ke and Qwen 3B student models, supporting absent audio as a default that requires no reference recording. RT-OPD outperforms ``Sharpening'' on all three benchmarks, suggesting that its gains are not solely due to sharpening the teacher distribution. Compared with the G-OPD-style ``Model reference'', RT-OPD yields higher Macro-3 for both students, suggesting that an audio-based contrast provides more effective supervision overall in these experiments. Finally, RT-OPD improves Macro-3 over ``Off-policy'' by 0.84 and 1.02 points for Ke and Qwen, respectively, supporting the benefit of on-policy over off-policy distillation.

\subsection{Audio Grounding Analysis}

\begin{table}[!t]

\centering

\caption{Audio perturbation on Ke-based 3B models (five-seed mean accuracy, \%). Parentheses show changes from the original audio (percentage points).}

\vspace{3pt}

\begin{tabular}{@{}lrrr@{}}

\toprule

Method & Orig. & Silence & Replaced\\

\midrule

\multicolumn{4}{l}{\emph{MMAU}}\\

Vanilla OPD & 72.21 & 52.20 ($-$20.01) & 48.56 ($-$23.65)\\

RT-OPD & 72.72 & 52.03 ($-$20.69) & 47.07 ($-$25.65)\\

\midrule

\multicolumn{4}{l}{\emph{MMAR}}\\

Vanilla OPD & 57.78 & 38.62 ($-$19.16) & 38.74 ($-$19.04)\\

RT-OPD & 60.08 & 37.62 ($-$22.46) & 38.44 ($-$21.64)\\

\midrule

\multicolumn{4}{l}{\emph{ADQA-cl.}}\\

Vanilla OPD & 56.02 & 24.27 ($-$31.74) & 26.32 ($-$29.70)\\

RT-OPD & 56.45 & 25.16 ($-$31.29) & 26.24 ($-$30.21)\\

\bottomrule

\end{tabular}

\label{tab:4}
\end{table}

To examine reliance on acoustic evidence, we evaluate the Ke-based models trained with Vanilla OPD and RT-OPD, each under three input conditions: original audio, silence, and replacement with another recording (Table~\ref{tab:4}). RT-OPD generally exhibits larger performance drops under audio perturbations, suggesting greater reliance on acoustic evidence.

\section{Conclusion}

This paper presents RT-OPD, which uses the teacher's audio-present/absent log-probability contrast to reshape the teacher distribution and strengthen acoustic grounding during on-policy distillation. Across two students and three benchmarks, RT-OPD consistently outperforms Vanilla OPD and the other trained methods in Macro-3, while audio perturbation analysis suggests stronger reliance of RT-OPD on acoustic evidence. Mizar-3B achieves the highest MMAU accuracy among the compared 3B models and remains competitive with several 7B and 8B models (Table~\ref{tab:1}).

\end{document}

%% file: figures/figure1_tikz.tex
\begin{tikzpicture}[
  x=1pt, y=1pt,
  font=\small,
  line width=0.6pt,
  line join=round,
  arr/.style={-{Stealth[length=4pt,width=3.2pt]}},
  graybox/.style={draw, rounded corners=3pt, fill=black!6, align=center, inner sep=2pt},
  whitebox/.style={draw, rounded corners=3pt, fill=white, align=center, inner sep=2pt},
  lbl/.style={inner sep=1pt},
  barM/.style={draw=black, line width=0.5pt, fill=black!45},
  barN/.style={draw=black, line width=0.5pt, fill=white},
  barQ/.style={draw=black, line width=0.5pt, fill=black},
  icon/.style={line width=0.8pt, line cap=round}
]

\draw[icon, domain=0:40, samples=161, variable=\x]
  plot (\x, {140+(1.5+7.5*exp(-(\x-20)*(\x-20)/80))*sin(80*\x)});
\node[lbl, align=center, anchor=north] at (20,128) {Matched\\audio $x^{M}$};

\draw[icon] (20,88) circle[radius=5.5pt];
\draw[icon] (15,82) -- (25,94);
\node[lbl, align=center, anchor=north] at (20,78) {No audio\\input $x^{\varnothing}$};

\node[graybox, minimum width=56pt, minimum height=28pt] (tT) at (88,140) {Frozen\\teacher};
\node[graybox, minimum width=56pt, minimum height=28pt] (tB) at (88,88) {Frozen\\teacher};
\draw[arr] (44,140) -- (tT.west);
\draw[arr] (44,88) -- (tB.west);
\draw[arr] (tT.east) -- (131,140);
\draw[arr] (tB.east) -- (131,88);

\foreach \gx/\pv in {142/0.22, 154/0.30, 166/0.26, 178/0.22}
  \draw[barM] (\gx-4,129) rectangle (\gx+4,{129+65*\pv});
\draw (134,129) -- (186,129);
\node[inner sep=2pt, anchor=south] at (160,150) {$p_{T,t}^{M}$};

\foreach \gx/\pv in {142/0.30, 154/0.15, 166/0.33, 178/0.22}
  \draw[barN] (\gx-4,77) rectangle (\gx+4,{77+65*\pv});
\draw (134,77) -- (186,77);
\node[inner sep=2pt, anchor=south] at (160,100) {$p_{T,t}^{\varnothing}$};

\node[whitebox, minimum width=58pt, minimum height=28pt] (cn) at (231,114) {Contrast\\+ normalize};
\draw[arr] (190,140) -| ([xshift=-17pt]cn.north);
\draw[arr] (190,88) -| ([xshift=-17pt]cn.south);
\draw[arr] (cn.east) -- (276,114);

\node[lbl, anchor=north, font=\small\bfseries] at (196,72) {Full-vocabulary contrast reward};
\node[lbl, anchor=north] at (196,60) {$r_t(v)=\log p_{T,t}^{M}(v)-\log p_{T,t}^{\varnothing}(v)$};

\draw[line width=0.8pt, rounded corners=5pt, fill=black!5] (276,60) rectangle (438,166);
\node[lbl, font=\small\bfseries] at (357,157) {Reward-tilted target};
\node[lbl] at (357,141) {$q_{\alpha,t}(v)\propto p_{T,t}^{M}(v)\exp(\alpha\,r_t(v))$};

\draw[barM] (309,120.5) rectangle (316,127.5);
\node[lbl, anchor=base west] at (317.5,122) {$p_{T,t}^{M}$};
\draw[barQ] (380,120.5) rectangle (387,127.5);
\node[lbl, anchor=base west] at (388.5,122) {$q_{\alpha,t}$};

\foreach \gx/\pv/\qv/\tok in {303.75/0.22/0.177/1, 339.25/0.30/0.399/2, 374.75/0.26/0.217/3, 410.25/0.22/0.207/4} {
  \draw[barM] (\gx-10,76) rectangle (\gx-1,{76+85*\pv});
  \draw[barQ] (\gx+1,76) rectangle (\gx+10,{76+85*\qv});
  \node[lbl, anchor=north] at (\gx,74) {$v_{\tok}$};
}
\draw (286,76) -- (428,76);

\node[graybox, minimum width=112pt, minimum height=20pt] (stu) at (56,22) {Student: matched audio};
\node[lbl, anchor=north] at (56,11) {Update student only};
\node[inner sep=2pt] (pS) at (172,22) {$p_{S,t}^{M}$};
\node[inner sep=2pt] (kl) at (282,22) {$D_{\mathrm{KL}}(p_{S,t}^{M}\,\|\,q_{\alpha,t})$};
\draw[arr] (stu.east) -- (pS.west);
\draw[arr] (pS.east) -- (kl.west);
\draw[arr] (357,60) |- (kl.east);

\end{tikzpicture}%